\documentclass[12pt]{article}
\newcommand{\nc}{\newcommand}
\nc{\renc}{\renewcommand}

\usepackage{graphicx}

\nc{\half}{{\textstyle{1\over2}}}
\nc{\etal}{\mbox{\it et al. }}
\nc{\ie}{{\it i.e.}}
\nc{\eg}{{\it e.g.}}

\renc{\thefootnote}{\arabic{footnote}}
\nc{\capt}[1]{{\bf Figure.} {\small\sl #1}}

\nc{\eqs}[2]{\mbox{Eqs.~(\ref{#1},\,\ref{#2})}}
\nc{\eq}[1]{\mbox{Eq.~(\ref{#1})}}

\nc{\figs}[2]{\mbox{Figs.~(\ref{#1},\,\ref{#2})}}
\nc{\fig}[1]{\mbox{Fig~.(\ref{#1})}}

\nc{\tag}[1]{\label{#1} \marginpar{{\footnotesize #1}}}
\nc{\mtag}[1]{\label{#1} \mbox{\marginpar{{\footnotesize #1}}}}
\renc{\baselinestretch}{1.5}
\newlength{\overeqskip}
\newlength{\undereqskip}
\nc{\be}[1]{\begin{equation} \mbox{$\label{#1}$}}
\nc{\bea}[1]{\begin{eqnarray} \mbox{$\label{#1}$}}
\nc{\Section}[2]{\section{#2}\label{#1}}
\nc{\Bibitem}[1]{\bibitem{#1}}
\nc{\Label}[1]{\label{#1}}

\nc{\eea}{\vspace{\undereqskip}\end{eqnarray}}
\nc{\ee}{\vspace{\undereqskip}\end{equation}}
\nc{\bdm}{\begin{displaymath}}
\nc{\edm}{\end{displaymath}}
\nc{\dpsty}{\displaystyle}
\nc{\bc}{\begin{center}}
\nc{\ec}{\end{center}}
\nc{\ba}{\begin{array}}
\nc{\ea}{\end{array}}
\nc{\bab}{\begin{abstract}}
\nc{\eab}{\end{abstract}}
\nc{\btab}{\begin{tabular}}
\nc{\etab}{\end{tabular}}
\nc{\bit}{\begin{itemize}}
\nc{\eit}{\end{itemize}}
\nc{\ben}{\begin{enumerate}}
\nc{\een}{\end{enumerate}}
\nc{\bfig}{\begin{figure}}
\nc{\efig}{\end{figure}}
\nc{\arreq}{&\!=\!&}
\nc{\arrmi}{&\!-\!&}
\nc{\arrpl}{&\!+\!&}
\nc{\arrap}{&\!\!\!\approx\!\!\!&}
\nc{\non}{\nonumber\\*}
\nc{\align}{\!\!\!\!\!\!\!\!&&}

\def\lsim{\; \raise0.3ex\hbox{$<$\kern-0.75em
      \raise-1.1ex\hbox{$\sim$}}\; }
\def\gsim{\; \raise0.3ex\hbox{$>$\kern-0.75em
      \raise-1.1ex\hbox{$\sim$}}\; }
\nc{\DOT}{\hspace{-0.08in}{\bf .}\hspace{0.1in}}
\nc{\Laada}{\hbox {$\sqcap$ \kern -1em $\sqcup$}}
\nc\loota{{\scriptstyle\sqcap\kern-0.55em\hbox{$\scriptstyle\sqcup$}}}
\nc\Loota{{\sqcap\kern-0.65em\hbox{$\sqcup$}}}
\nc\laada{\Loota}
\nc{\qed}{\hskip 3em \hbox{\BOX} \vskip 2ex}

\nc{\real}{{\rm I \! R}}
\nc{\Z}{{\sf Z \!\!\! Z}}
\nc{\complex}{{\rm C\!\!\! {\sf I}\,\,}}
\def\bigid{\leavevmode\hbox{\small1\kern-3.8pt\normalsize1}}
\def\id{\leavevmode\hbox{\small1\kern-3.3pt\normalsize1}}
\nc{\slask}{\!\!\!/}
\nc{\bis}{{\prime\prime}}
\nc{\pa}{\partial}
\nc{\na}{\nabla}
\nc{\ra}{\rangle}
\nc{\la}{\langle}
\nc{\goto}{\rightarrow}
\nc{\swap}{\leftrightarrow}

\nc{\EE}[1]{ \mbox{$\cdot10^{#1}$} }
\nc{\abs}[1]{\left|#1\right|}
\nc{\at}[2]{\left.#1\right|_{#2}}
\nc{\norm}[1]{\|#1\|}
\nc{\abscut}[2]{\Abs{#1}_{\scriptscriptstyle#2}}
\nc{\vek}[1]{{\rm\bf #1}}
\nc{\integral}[2]{\int\limits_{#1}^{#2}}
\nc{\inv}[1]{\frac{1}{#1}}
\nc{\dd}[2]{{{\partial #1}\over{\partial #2}}}
\nc{\ddd}[2]{{{{\partial}^2 #1}\over{\partial {#2}^2}}}
\nc{\dddd}[3]{{{{\partial}^2 #1}\over
        {\partial #2 \partial #3}}}
\nc{\dder}[2]{{{d #1}\over{d #2}}}
\nc{\ddder}[2]{{{d^2 #1}\over{d {#2}^2}}}
\nc{\dddder}[3]{{d^2 #1}\over
        {d #2 d #3}}
\nc{\dx}[1]{d\,^{#1}x}
\nc{\dy}[1]{d\,^{#1}y}
\nc{\dz}[1]{d\,^{#1}z}
\nc{\dl}[1]{\frac{d\,^{#1}l}{(2\pi)^{#1}}}
\nc{\dk}[1]{\frac{d\,^{#1}k}{(2\pi)^{#1}}}
\nc{\dq}[1]{\frac{d\,^{#1}q}{(2\pi)^{#1}}}

\nc{\cc}{\mbox{$c.c.$ }}
\nc{\hc}{\mbox{$h.c.$ }}
\nc{\cf}{cf.\ }
\nc{\erfc}{{\rm erfc}}
\nc{\Tr}{{\rm Tr\,}}
\nc{\tr}{{\rm tr\,}}
\nc{\pol}{{\rm pol}}
\nc{\sign}{{\rm sign}}
\nc{\bfT}{{\bf T }}

\nc{\cA}{{\cal A}}
\nc{\cB}{{\cal B}}
\nc{\cD}{{\cal D}}
\nc{\cE}{{\cal E}}
\nc{\cG}{{\cal G}}
\nc{\cH}{{\cal H}}
\nc{\cL}{{\cal L}}
\nc{\cO}{{\cal O}}
\nc{\cT}{{\cal T}}
\nc{\cN}{{\cal N}}
\nc{\rvac}[1]{|{\cal O}#1\rangle}
\nc{\lvac}[1]{\langle{\cal O}#1|}
\nc{\rvacb}[1]{|{\cal O}_\beta #1\rangle}
\nc{\lvacb}[1]{\langle{\cal O}_\beta #1 |}
\nc{\bb}{\bar{\beta}}
\nc{\bt}{\tilde{\beta}}
\nc{\ctH}{\tilde{\cal H}}
\nc{\chH}{\hat{\cal H}}
\nc{\1}{\aa}
\nc{\2}{\"{a}}
\nc{\3}{\"{o}}
\nc{\4}{\AA}
\nc{\5}{\"{A}}
\nc{\6}{\"{O}}
\nc{\al}{\alpha}
\nc{\g}{\gamma}
\nc{\Del}{\Delta}
\nc{\e}{\epsilon}
\nc{\eps}{\epsilon}
\nc{\lam}{\lambda}
\nc{\om}{\omega}
\nc{\Om}{\Omega}
\nc{\ve}{\varepsilon}
\nc{\mn}{{\mu\nu}}
\nc{\vp}{\varphi}

\nc{\advp}[3]{{\it  Adv.\ in\ Phys.\ }{{\bf #1} {(#2)} {#3}}}
\nc{\annp}[3]{{\it  Ann.\ Phys.\ (N.Y.)\ }{{\bf #1} {(#2)} {#3}}}
\nc{\apl}[3]{{\it  Appl. Phys. Lett. }{{\bf #1} {(#2)} {#3}}}
\nc{\apj}[3]{{\it  Ap.\ J.\ }{{\bf #1} {(#2)} {#3}}}
\nc{\apjl}[3]{{\it  Ap.\ J.\ Lett.\ }{{\bf #1} {(#2)} {#3}}}
\nc{\app}[3]{{\it Astropart.\ Phys.\ }{{\bf #1} {(#2)} {#3}}}
\nc{\cmp}[3]{{\it  Comm.\ Math.\ Phys.\ }{{ \bf #1} {(#2)} {#3}}}
\nc{\cqg}[3]{{\it  Class.\ Quant.\ Grav.\ }{{\bf #1} {(#2)} {#3}}}
\nc{\epl}[3]{{\it  Europhys.\ Lett.\ }{{\bf #1} {(#2)} {#3}}}
\nc{\ijmp}[3]{{\it Int.\ J.\ Mod.\ Phys.\ }{{\bf #1} {(#2)} {#3}}}
\nc{\ijtp}[3]{{\it Int.\ J.\ Theor.\ Phys.\ }{{\bf #1} {(#2)} {#3}}}
\nc{\jmp}[3]{{\it  J.\ Math.\ Phys.\ }{{ \bf #1} {(#2)} {#3}}}
\nc{\jpa}[3]{{\it  J.\ Phys.\ A\ }{{\bf #1} {(#2)} {#3}}}
\nc{\jpc}[3]{{\it  J.\ Phys.\ C\ }{{\bf #1} {(#2)} {#3}}}
\nc{\jap}[3]{{\it J.\ Appl.\ Phys.\ }{{\bf #1} {(#2)} {#3}}}
\nc{\jpsj}[3]{{\it J.\ Phys.\ Soc.\ Japan\ }{{\bf #1} {(#2)} {#3}}}
\nc{\lmp}[3]{{\it Lett.\ Math.\ Phys.\ }{{\bf #1} {(#2)} {#3}}}
\nc{\mpl}[3]{{\it  Mod.\ Phys.\ Lett.\ }{{\bf #1} {(#2)} {#3}}}
\nc{\ncim}[3]{{\it  Nuov.\ Cim.\ }{{\bf #1} {(#2)} {#3}}}
\nc{\np}[3]{{\it  Nucl.\ Phys.\ }{{\bf #1} {(#2)} {#3}}}
\nc{\npps}[3]{{\it  Nucl.\ Phys.\ Proc.\ Suppl.\ }{{\bf #1} {(#2)} {#3}}}
\nc{\pr}[3]{{\it Phys.\ Rev.\ }{{\bf #1} {(#2)} {#3}}}
\nc{\pra}[3]{{\it  Phys.\ Rev.\ A\ }{{\bf #1} {(#2)} {#3}}}
\nc{\prb}[3]{{\it  Phys.\ Rev.\ B\ }{{{\bf #1} {(#2)} {#3}}}}
\nc{\prc}[3]{{\it  Phys.\ Rev.\ C\ }{{\bf #1} {(#2)} {#3}}}
\nc{\prd}[3]{{\it  Phys.\ Rev.\ D\ }{{\bf #1} {(#2)} {#3}}}
\nc{\prl}[3]{{\it Phys.\ Rev.\ Lett.\ }{{\bf #1} {(#2)} {#3}}}
\nc{\pl}[3]{{\it  Phys.\ Lett.\ }{{\bf #1} {(#2)} {#3}}}
\nc{\prep}[3]{{\it Phys.\ Rep.\ }{{\bf #1} {(#2)} {#3}}}
\nc{\prsl}[3]{{\it Proc.\ R.\ Soc.\ London\ }{{\bf #1} {(#2)} {#3}}}
\nc{\ptp}[3]{{\it  Prog.\ Theor.\ Phys.\ }{{\bf #1} {(#2)} {#3}}}
\nc{\ptps}[3]{{\it  Prog\ Theor.\ Phys.\ suppl.\ }{{\bf #1} {(#2)} {#3}}}
\nc{\physa}[3]{{\it  Physica\ A\ }{{\bf #1} {(#2)} {#3}}}
\nc{\physb}[3]{{\it  Physica\ B\ }{{\bf #1} {(#2)} {#3}}}
\nc{\phys}[3]{{\it Physica\ }{{\bf #1} {(#2)} {#3}}}
\nc{\rmp}[3]{{\it  Rev.\ Mod.\ Phys.\ }{{\bf #1} {(#2)} {#3}}}
\nc{\rpp}[3]{{\it Rep.\ Prog.\ Phys.\ }{{\bf #1} {(#2)} {#3}}}
\nc{\sjnp}[3]{{\it Sov.\ J.\ Nucl.\ Phys.\ }{{\bf #1} {(#2)} {#3}}}
\nc{\spjetp}[3]{{\it Sov.\ Phys.\ JETP\ }{{\bf #1} {(#2)} {#3}}}
\nc{\yf}[3]{{\it Yad.\ Fiz.\ }{{\bf #1} {(#2)} {#3}}}
\nc{\zetp}[3]{{\it Zh.\ Eksp.\ Teor.\ Fiz.\  }{{\bf #1}  {(#2)} {#3}}}
\nc{\zp}[3]{{\it Z.\ Phys.\ }{{\bf #1} {(#2)} {#3}}}
\nc{\ibid}[3]{{\sl ibid.\ }{{\bf #1} {#2} {#3}}}
\nc{\rf}[1]{(\ref{#1})}
\nc{\nn}{\nonumber \\*}
\nc{\bfB}{\bf{B}}
\nc{\bfv}{\bf{v}}
\nc{\bfx}{\bf{x}}
\nc{\bfy}{\bf{y}}
\nc{\vx}{\vec{x}}
\nc{\vy}{\vec{y}}
\nc{\oB}{\overline{B}}
\nc{\oI}{\overline{I}}
\nc{\oR}{\overline{R}}
\nc{\rar}{\rightarrow}
\nc{\ti}{\times}
\nc{\slsh}{\hskip-5pt/}
\nc{\sm}{Standard~Model~}
\nc{\MP}{M_{\rm Pl}}
\nc{\tp}{t_{\rm Pl}}
\nc{\ave}{\bar{E}}

\nc{\eff}{{\rm eff}}
\nc{\kk}{\vek{k}}
\nc{\pp}{{\rm p}}
\nc{\ga}{g_{a\gamma}}
\nc{\vv}{\\}
\nc{\eee}{{\bf E}}
\nc{\bbb}{{\bf B}}
\nc{\qcd}{T_{\rm QCD}}
\nc{\G}{\rm \ G}
\def\vec#1{{\bf #1}}

\def\ell{e^{c}LL}

\begin{document}
{\title{\vskip-2truecm{\hfill {{\small \\
	\hfill \\
	}}\vskip 1truecm}
{\bf  Two-Field Q-ball Solutions of Supersymmetric Hybrid Inflation}}
{\author{
{\sc  Matt Broadhead$^{1}$ and John McDonald$^{2}$}\\
{\sl\small Dept. of Mathematical Sciences, University of Liverpool,
Liverpool L69 3BX, England}
}
\maketitle
\begin{abstract}
\noindent

        We demonstrate the existence of 
two-field Q-ball solutions of the scalar field equations of 
supersymmetric D- and F-term hybrid inflation. The solutions consist of a complex
 inflaton field together with a real symmetry  breaking field. Such inflatonic
 Q-balls may play a fundamental role in reheating and the post-inflation era of
 supersymmetric hybrid inflation models.

\end{abstract}
\vfil
\footnoterule
{\small $^1$mattb@amtp.liv.ac.uk}
{\small $^2$mcdonald@amtp.liv.ac.uk}

\thispagestyle{empty}
\newpage
\setcounter{page}{1}

\section{Introduction}

            Hybrid inflation models \cite{hi,lr} are a
 favoured class of inflation model, being able to account 
for both a flat inflaton potential during inflation and for a
 massive inflaton and reheating after
 inflation, without requiring very small couplings.
 In supersymmetry (SUSY) there are two
 classes of hybrid inflation model, D-term inflation \cite{dti} and 
F-term inflation \cite{fti}, depending on
 whether the energy density driving inflation originates in a D-term or F-term
 contribution to the scalar potential. 

               An important period in early Universe cosmology is the era
immediately following the end of inflation, 
the post-inflation era. Important physical processes such as baryogenesis 
are likely to 
occur during the post-inflation era, whilst  
reheating of the Universe will occur at the end of this era.
In hybrid inflation models it is known that quantum fluctuations of the inflaton
 sector fields will rapidly grow and become non-linear at the end of inflation
 \cite{tp,icf,cpr,matt1}. 
The question of the subsequent 
evolution of the non-linear field configurations in realistic SUSY inflation models 
remains to be fully explored, but one possibility is that
 non-topological soliton
 configurations will form \cite{icf,cpr,iballs}. The most stable such configuration
 will tend to be the 
dominant one, since this will come to dominate the energy density of the Universe
 and so determine the physics of the post-inflation era.

              In SUSY hybrid inflation models the scalar
 fields are generally complex, and therefore can carry conserved global
 $U(1)$ charges. Depending on the form of the scalar
 potential, it is then possible that a Q-ball made of 
inflaton sector fields exists \cite{qb}.  If such Q-balls formed at the end of inflation, 
the Universe following inflation would be highly 
inhomogeneous, with all the energy density concentrated 
in the form of inflatonic
 Q-balls. Post-inflation physics would then take place
 against this cosmological background \cite{icf,matt1,fdd}, whilst
reheating would occur via the eventual decay of the Q-balls\footnote{This 
possibility has previously been realised in the 
context of a single field chaotic inflation model \cite{ke1}.}.

     The purpose of this paper is to demonstrate the existence of inflatonic Q-balls 
in SUSY hybrid inflation
 models. We will present numerical 
examples of two-field Q-ball solutions, composed of
a complex inflaton field carrying a global
 $U(1)$ charge and a real symmetry breaking field.

        The paper is organised as follows. 
In Section 2 we discuss SUSY hybrid inflation models and the Q-ball equations. 
In Section 3 we demonstrate the existence of two field Q-ball solutions of these
 equations for the case of D- and F-term inflation. 
 In Section 4 we present our conclusions. 

\section{Q-ball Equations of SUSY Hybrid Inflation} 
               
              SUSY hybrid inflation models are either F-term or D-term models. 
The simplest F-term inflation model has a superpotential of the form \cite{lr,fti}
\be{hi1} W = \frac{\eta}{2} S(\Phi^{2} - \mu^{2})      ~,\ee 
where $S$ is the inflaton and $\Phi$ is a field which gains an expectation
 value which terminates inflation. $\mu^{2}$ and $\eta$ are real and positive. 
 The scalar potential is then
\be{hi2} V =   \eta^{2} |S|^{2}|\Phi|^{2}  + 
  \frac{\eta^{2}}{4} |\Phi^{2}-\mu^{2}|^{2}     
~.\ee
The scalar potential has an R-symmetry under which only $S$ transforms, which 
manifests itself as a global $U(1)$ symmetry in the scalar potential, $S \rightarrow 
e^{i \alpha}S$, with respect to which we can define a conserved global charge. 

           D-term inflation models have a superpotential of the form \cite{dti}
\be{e1}    W = \lambda S \Phi_{+} \Phi_{-}    ~.\ee
The scalar potential is given by,
\be{e2} V = \lambda^{2}|S|^{2}\left(|\Phi_{+}|^{2} + |\Phi_{-}|^{2}\right) 
+ \lambda^{2} |\Phi_{+}|^{2} |\Phi_{-}|^{2} + \frac{g^{2}}{2} \left(|\Phi_{+}|^{2}
- |\Phi_{-}|^{2} + \xi\right)^{2}    ~,\ee
where $S$ is the inflaton, $\Phi_{\pm}$ are fields with charges $\pm$1 with respect
 to a 
Fayet-Illiopoulos $U(1)_{FI}$ gauge symmetry, $\xi > 0$ is the Fayet-Illiopoulos
 term and 
$g$ is the $U(1)_{FI}$ gauge coupling. 

   The D-term inflation scalar potential is a function 
of $|S|$, $|\Phi_{+}|$ and $|\Phi_{-}|$ and therefore 
has three global $U(1)$ symmetries, $S \rightarrow e^{i \alpha} S$ ($U(1)_{S}$),
$\Phi_{+} \rightarrow e^{i \beta_{+}} \Phi_{+}$ ($U(1)_{+}$)
and $\Phi_{-} \rightarrow e^{i \beta_{-}} \Phi_{-}$ ($U(1)_{-}$). 
We can define conserved charges with
respect to these global $U(1)$ symmetries, $Q_{S}$, $Q_{+}$ and $Q_{-}$. 

           Since SUSY hybrid inflation models have conserved global charges, 
it is possible that there exist Q-balls \cite{qb}.
We first review the case of a single complex field $\Phi$ with $U(1)$ symmetry 
$\Phi \rightarrow e^{i \alpha} \Phi$. The Q-ball configuration is
 derived by minimizing the energy whilst fixing
 the charge via a Lagrange multiplier i.e. 
by minimizing the functional \cite{ks}
\be{e3} E_{\omega}= E + \omega \left(Q - \int d^{3}x \rho_{Q}\right)   ~,\ee
with respect to the scalar fields and $\omega$, where $E$ is the total energy of the
 field configuration
\be{e4} E = \int d^{3}x \; |\dot{\Phi}|^{2} + |\underline{\nabla} \Phi|^{2}
+ V(|\Phi|)    ~\ee 
and $\rho_{Q}$ is the charge density
\be{e5} \rho_{Q} = i( \dot{\Phi}^{\dagger}\Phi - \Phi^{\dagger} \dot{\Phi})  ~.\ee
$E_{\omega}$ may be equivalently written as 
\be{e6} E_{\omega}(\dot{\Phi},\Phi,\omega)  = \int d^{3}x \left( |\dot{\Phi}
 - i \omega \Phi|^{2} + 
 |\underline{\nabla}\Phi|^{2} + V(\Phi) - \omega^{2} |\Phi|^{2} \right) + \omega Q  
 ~.\ee
This should be minimized with respect to $\dot{\Phi}$, $\Phi$ and $\omega$.
To minimize with respect to $\dot{\Phi}$ we require $\Phi(\vec{x},t)
 = \Phi(\vec{x}) e^{i \omega t}$. Substituting this into \eq{e6} gives 
\be{e7} E_{\omega}(\Phi(\vec{x}),\omega) = \int d^{3}x \left(
 |\underline{\nabla}\Phi(\vec{x})|^{2}
 + V(\Phi(\vec{x})) - \omega^{2} |\Phi(\vec{x})|^{2} \right) + \omega Q  ~.\ee
Extremizing this with respect to $\Phi(\vec{x})$ implies that
\be{e8}  \underline{\nabla}^{2} \Phi(\vec{x}) = \frac{\partial
 V_{\omega}\left(\Phi(\vec{x})\right)}{\partial \Phi^{\dagger}}   ~\ee
where $V_{\omega} = V - \omega^{2} |\Phi|^{2}$.   
At this point $\Phi(\vec{x})$ could still have a space-dependent complex phase, 
$\theta(\vec{x})$. 
If $V(\Phi) = V(|\Phi|)$, as it must when $\Phi$ transforms under a $U(1)$ symmetry, 
then \eq{e7} is generally
 minimized by the choice $\theta = $ {\it constant}, 
which may be chosen such that $\Phi(\vec{x})$ is real. 
A minimum energy configuration should be spherically symmetric. 
Then, with $\Phi(\vec{x}) = \phi(r)/\sqrt{2}$ ($\phi(r)$ real), \eq{e8} becomes 
\be{e9}  \frac{ \partial^{2} \phi}{\partial r^{2}}  +\frac{2}{r} \frac{\partial
 \phi}{\partial r} = \frac{ \partial V}{\partial \phi} - \omega^{2} \phi    ~.\ee
We refer to this as the Q-ball equation. 
The solutions of \eq{e9} should satisfy the boundary 
conditions that the field tends to the vacuum as 
$r \rightarrow \infty$ and that $\partial \phi/\partial 
r \rightarrow 0$ as $r \rightarrow 0$.

     The above analysis generalizes to the multiple complex 
scalar field case of SUSY hybrid inflation. 
For each value of $\omega$ there can be many 
solutions of the Q-ball equations satisfying the boundary conditions, 
each with a different energy and charge. In particular, there will be a 
range of solutions of different energy and $\omega$ for a given global 
charge. 
Each of these solutions is a Q-ball in the sense that it is 
a solution of the scalar field equations corresponding to a
non-topological soliton which has  
a time-independent amplitude as a function of $r$. 
However, these Q-balls will be metastable with respect to
the lowest energy Q-ball solution. 
The stable Q-ball solution is the lowest energy 
field configuration for a given global charge, 
obtained by minimizing the energy functional 
with respect to $\omega$ for a fixed charge. 

      We will refer to solutions of \eq{e9} which are not minimum energy 
solutions for a given charge as 'metastable Q-balls'. 
 The existence of one metastable Q-ball solution for a given charge is sufficient
 to prove the existence of the stable Q-ball; it is either the minimum energy 
solution itself or there exists a lower energy solution of
 \eq{e9} carrying the same global charge.

             In the following we will focus on the case of D-term inflation.
This is because, as we will show, the 
Q-ball equations for the case of F-term inflation
are equivalent to those of D-term inflation with $\lambda = \sqrt{2}g$. Therefore
the F-term inflation Q-balls are a subset of those of D-term inflation.

     We now consider three Lagrange multipliers 
$\omega$, $\gamma_{+}$ and $\gamma_{-}$, corresponding to the conserved 
charges $Q_{S}$, $Q_{+}$ and $Q_{-}$ respectively. The functional is now 
\be{e10}  E_{\omega} = E + \omega \left(Q_{S} - \int d^{3}x \rho_{Q_{S}}\right) 
+ \gamma_{+} \left(Q_{+} - \int d^{3}x \rho_{Q_{+}}\right)  
+ \gamma_{-} \left(Q_{-} - \int d^{3}x \rho_{Q_{-}}\right)
~,\ee 
where 
\be{e11} E = \int d^{3}x \;  |\dot{S}|^{2} + |\underline{\nabla} S|^{2} +
|\dot{\Phi}_{+}|^{2} + |\underline{\nabla} \Phi_{+}|^{2} +
|\dot{\Phi}_{-}|^{2} + |\underline{\nabla} \Phi_{-}|^{2}
+ V(|S|, |\Phi_{+}|, |\Phi_{-}|)    ~,\ee 
\be{e12} \rho_{Q_{S}} = i( \dot{S}^{\dagger}S - S^{\dagger} \dot{S})  ~\ee
and
\be{e13} \rho_{Q_{\pm}} = i( \dot{\Phi}_{\pm}^{\dagger}
\Phi_{\pm} - \Phi_{\pm}^{\dagger}
 \dot{\Phi}_{\pm})  ~.\ee
As before, minimizing the time derivative terms implies that 
$S(\vec{x},t )= S(\vec{x})e^{i \omega t}$ and $\Phi_{\pm}(\vec{x},t) =
 \Phi_{\pm}(\vec{x})e^{i \gamma_{\pm} t}$. 
For D-term inflation, $V = V(|S|,|\Phi_{+}|,|\Phi_{-}|)$. Therefore the minimum
 energy 
configuration will correspond to real $S(\vec{x})$ and $\Phi_{\pm}(\vec{x})$. 
Assuming a spherically symmetric minimum energy configuration 
then implies that 
$S = s(r)e^{i \omega t}/\sqrt{2}$ and $\Phi_{\pm} =
 \phi_{\pm}(r)e^{i\gamma_{\pm} t}/\sqrt{2}$. The vacuum of D-term inflation 
corresponds to $|\Phi_{-}| = \xi^{1/2}$ and $S = \Phi_{+} = 0$. Therefore we must
 have $\gamma_{-} = 0$
in order to reach the $\Phi_{-}$ vacuum expectation value as $r \rightarrow \infty$. 
Thus the D-term inflation Q-ball must have $Q_{-} = 0$. 
Since $\Phi_{+} \rightarrow 0$ as $r \rightarrow \infty$, the Q-ball could, in
 principle, 
carry a $U(1)_{+}$ charge. However, it is unlikely that a Q-ball solution with a
 $U(1)_{+}$ 
charge exists. This is because the effective mass of the $\Phi_{+}$ scalar increases 
as $|\Phi_{-}|$ decreases from $\xi$ and $S$ increases from zero as $r \rightarrow 0$. 
Thus a $U(1)_{+}$ charge is likely to be energetically disfavoured. 
Therefore we will focus on the case $Q_{+} = 0$, for which $\gamma_{+} = 0$.

              The corresponding Q-ball equations are then
\be{e14}  \frac{ \partial^{2} s}{\partial r^{2}}  +\frac{2}{r} \frac{\partial s}{\partial
 r} = \frac{\lambda^{2}}{2}
 \left(\phi_{+}^{2} +\phi_{-}^{2}\right) s - \omega^{2} s    ~,\ee
\be{e15a}  \frac{ \partial^{2} \phi_{+}}{\partial r^{2}}  +\frac{2}{r} \frac{\partial
 \phi_{+}}{\partial r} =  
\frac{\lambda^{2}}{2}\left(s^{2}+\phi_{-}^{2}\right)\phi_{+}  
+ g^{2} \left(\xi - \frac{\phi_{-}^{2}}{2}\right)\phi_{+} + 
\frac{g^{2}}{2} \phi_{+}^{3} 
 ~\ee     
and
\be{e15b}  \frac{ \partial^{2} \phi_{-}}{\partial r^{2}}  +\frac{2}{r} \frac{\partial
 \phi_{-}}{\partial r} =  
\frac{\lambda^{2}}{2}\left(s^{2}+\phi_{+}^{2}\right)\phi_{-}  
- g^{2} \left(\xi + \frac{\phi_{+}^{2}}{2}\right)\phi_{-} + 
\frac{g^{2}}{2} \phi_{-}^{3} 
 ~.\ee     
For $\phi_{-}^{2} \leq 2 \xi$, which will be true for any 
 solution tending to the vacuum as $r \rightarrow \infty$, 
the only solution of \eq{e15a} which satisfies the boundary conditions $\phi_{+}
 \rightarrow 0$ 
as $r \rightarrow \infty$ and $\partial \phi_{+}/\partial r \rightarrow 0$ as $r \rightarrow 0$ 
is $\phi_{+}(r) = 0$ $\forall r$. This follows since for a minimum energy solution 
we expect 
$\partial \phi_{+}/\partial r \leq 0$, such that $\phi_{+}(r)$ is monotonically decreasing 
to zero as $r$ increases. Since the right hand side of \eq{e15a} is positive $\forall r$, 
it then follows that $\partial^{2} \phi_{+} /\partial r^{2} > 0$ $\forall r$. However, 
for a monotonically decreasing $\phi_{+}(r)$ we require that 
$\partial^{2} \phi_{+} /\partial r^{2} < 0$ at $r=0$. 
Therefore there is no non-trivial solution. Thus 
for a Q-ball with $Q_{S} \neq 0$ and $Q_{+} = 0$, the Q-ball equations become 
\be{e16}  \frac{ \partial^{2} s}{\partial r^{2}}  +\frac{2}{r} \frac{\partial s}{\partial
 r} = \frac{\lambda^{2}}{2}\phi_{-}^{2} s - \omega^{2} s    ~\ee
and
\be{e17}  \frac{ \partial^{2} \phi_{-}}{\partial r^{2}}  +\frac{2}{r} \frac{\partial
 \phi_{-}}{\partial r} = 
\left(\frac{\lambda^{2}}{2}s^{2}  - g^{2} \xi\right)\phi_{-} + \frac{g^{2}}{2} \phi_{-}^{3} 
 ~.\ee     
Therefore the Q-ball solution of D-term inflation with $Q_{+} = 0$ 
consists of a complex $S$ field and a real
$\Phi_{-}$ field, with a $Q_{S}$ charge but no $Q_{-}$ charge. 
The energy and charge of the resulting Q-ball soultion are given by
\be{e18}  E = \int 4 \pi r^{2} dr \left[ \frac{1}{2} \left( \frac{\partial s}{\partial
 r}\right)^{2} +
\frac{1}{2} \left( \frac{\partial \phi_{-}}{\partial r}\right)^{2} 
+ \frac{\omega^{2} s^{2}}{2} + V(s,\phi_{-}) \right] 
~\ee
and 
\be{e19} Q_{S} = \omega \int 4 \pi r^{2} dr \; s^{2} ~.\ee

    In the case of F-term inflation, only the $S$ field can carry a global charge. 
Therefore, upon performing the minimization of the energy
 functional, from minimizing 
with respect to $\dot{S}$ and $\dot{\Phi}$ we
 obtain $S(\vec{x},t) = S(\vec{x})e^{i \omega t}$ 
and $\Phi(\vec{x},t) = \Phi(\vec{x})$. 
The potential, \eq{hi2}, is explicitly dependent upon the phase of $\Phi$. 
However, for $\mu^{2}$ real and positive the energy functional will still 
be minimized by having both $S(\vec{x})$ and $\Phi(\vec{x})$ real and positive. 
Therefore with $S(\vec{x}) = s(r)/\sqrt{2}$ and $\Phi(\vec{x}) = \phi(r)/\sqrt{2}$, 
the Q-ball equations become
\be{e20}  \frac{ \partial^{2} s}{\partial r^{2}}  +\frac{2}{r} \frac{\partial s}{\partial
 r} = \frac{\eta^{2}}{2}\phi^{2} s - \omega^{2} s    ~\ee
and
\be{e21}  \frac{ \partial^{2} \phi}{\partial r^{2}}  +\frac{2}{r} \frac{\partial
 \phi}{\partial r} = 
\left(\frac{\eta^{2}}{2}s^{2}  - \frac{\eta^{2}}{2} \xi\right)\phi 
+ \frac{\eta^{2}}{4} \phi^{3} 
 ~.\ee       
These equations are the same as the two-field D-term inflation equations, 
\eq{e16} and \eq{e17}, when $\phi_{-} \leftrightarrow \phi$, $\lambda
 \leftrightarrow \eta$ 
and $g \leftrightarrow \eta/\sqrt{2}$ i.e. when $\lambda = \sqrt{2}g$.  

\section{Numerical Two-Field Q-Ball Solutions}

              In this section we will present a number of numerical solutions of 
\eq{e16} and \eq{e17} corresponding to Q-balls. We consider $g = 1$ throughout. 

             For a given $\lambda$ 
there will be a range of solutions corresponding to different values of $Q_{S}$. 
In Figure 1 we show a Q-ball solution for $s(r)$ 
and $\phi_{-}(r)$ for the case
$\lambda = 0.5$. (We use units such that $\xi = 1$.) 
In Figure 2 we show a solution for 
$\lambda = 1$. In Figure 3 we show a solution for the special case 
$\lambda = \sqrt{2}g$, 
corresponding to the case of F-term inflation. In Table 1 we summarise the properties 
of these example Q-ball solutions. (We define the radius as the value of $r$ within
 which $90\%$ of the total Q-ball energy is contained.) $E/Q_{S}$ is less than
 the $S$ mass in vacuum for all of these solutions
($m_{S}= \lambda \xi^{1/2} \equiv \lambda$ in our 
units), so in the absence of additional couplings to the MSSM fields
 the Q-balls will be absolutely stable as a result of  $Q_{S}$ conservation. 
(However, once the (unknown) couplings of the inflaton sector fields to 
the MSSM fields are included, the inflatonic Q-balls will decay to 
MSSM fields via conventional inflaton decay.)
An interesting feature of these solutions is that the value of 
$s(r = 0)$ at the centre of the 
Q-balls is {\it larger} than the value at which the symmetry breaking phase
 transition ending inflation occurs,  
$s(r = 0) > s_{c} = \sqrt{2} g \xi^{1/2}/\lambda$ ($\equiv 1.41 g/\lambda$ for
 $\xi = 1$). However, in the Q-ball solution, where the field 
configuration is dependent
 upon the gradient energy as well as the potential energy, the symmetry
 breaking field, $\phi_{-}$, remains non-zero throughout the Q-ball.

\begin{figure}[htbp]
\begin{center}
\includegraphics[width=0.75\textwidth]{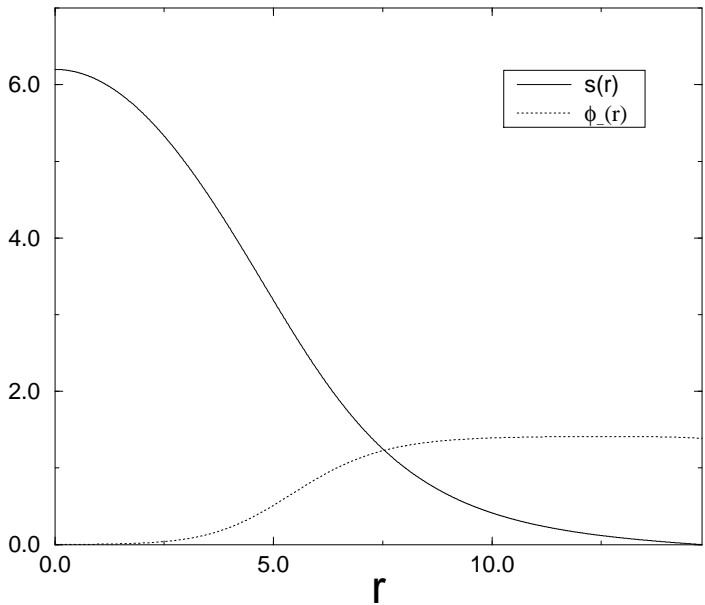}
\caption{\footnotesize{Q-ball profile for $\lambda=0.5$ and
$g = 1$.}}
\label{fig:fig2}
\end{center}
\end{figure}

\begin{table}[htbp]
\begin{center}
\begin{tabular}{|c|c|c|c|c|c|c|c|}
	\hline  $\lambda$ & $\omega$ & $s_{o}$
                     & $\phi_{o}$ & $E$ & $Q_{S}$ & $E/Q_{S}$ & $r$ \\
	\hline  $0.5$ & $0.375$ & $0.20$ &
 $0.00371$ & $2894$
		     & $6346$ & $0.456$ & $8.31$ \\
	\hline  $1.0$ & $0.800$ & $3.00$ & $0.14357$ & $378$ &
 $396$ & $0.956$ & $4.33$ \\
	\hline  $\sqrt{2}$ & $1.125$ & $2.44$ & $0.19680$ & $180$
 & $132$ & $1.368$ & $3.11$ \\
		\hline     
\end{tabular}
\caption{\footnotesize{Properties of example Q-ball solutions for $g=1$.}}  
\label{fig:fig1}
\end{center}
\end{table}

\newpage

       In Table 2 we give the values of $\omega$, $s(r=0)$, 
$\phi_{-}(r=0)$, $r$ and $E/Q_{S}$ for metastable Q-balls with $\lambda = 
1$, $g = 1$ and fixed charge $Q_{S} \approx 395$.
Since the value of $Q_{S}$ is generally much larger than 1, the Q-balls may be 
studied classically. 
Metastable Q-ball solutions exist for a finite range of $\omega$.
The lowest value of $E/Q_{S}$ corresponds to the true Q-ball solution. 
As $\omega$ decreases, the value of $E/Q_{S}$ decreases, 
implying that the binding energy of the $S$ charges in the Q-ball is
 increasing. An interesting feature is that the value of $s$ at the centre 
of the Q-ball increases whilst $r$ decreases as 
$E/Q_{S}$ increases, indicating that the metastable Q-balls have a 
larger gradient energy than the true Q-balls for a given charge.

\begin{figure}[hp]
\begin{center}
\includegraphics[width=0.75\textwidth]{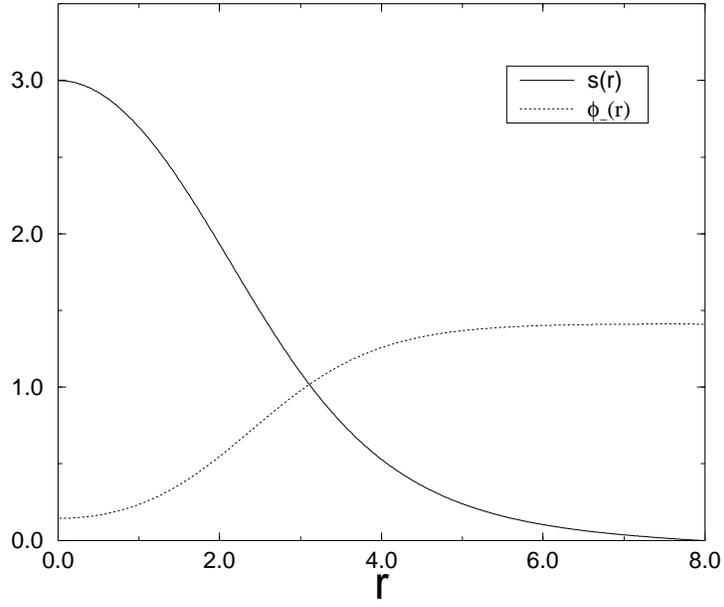}
\caption{\footnotesize{Q-ball profile for $\lambda = 1$ and
$g = 1$.}}
\label{fig:fig3}
\end{center}
\end{figure}

\begin{figure}[hp]
\begin{center}
\includegraphics[width=0.75\textwidth]{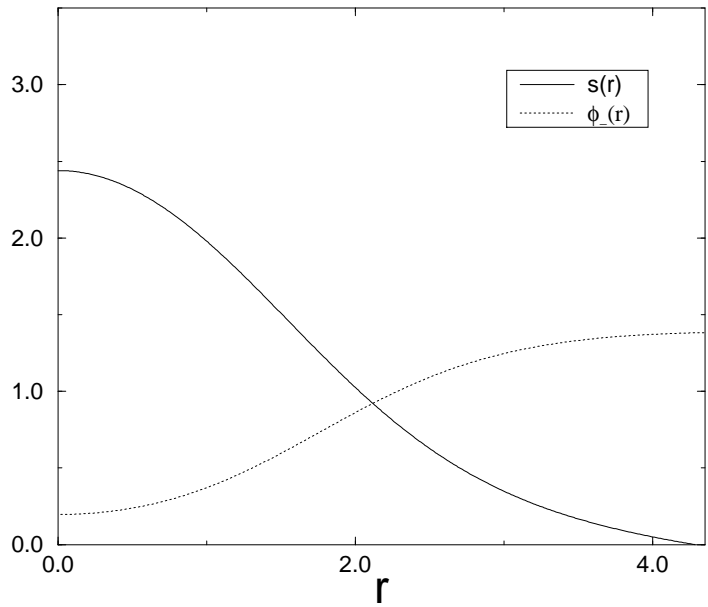}
\caption{\footnotesize{Q-ball profile 
for $\lambda =\sqrt{2}$ and
$g = 1$ (F-term inflation).}}
\label{fig:fig4}
\end{center}
\end{figure}

\begin{table}[htbp]
\begin{center}
\begin{tabular}{|c|c|c|c|c|}
	\hline  \bf{$\omega$} & $s_{o}$ & $\phi_{o}$ & $r$ 
& $E/Q_{S}$ \\
	\hline  $0.800$ & $3.00$ & $0.1436$ & $4.33$ & 0.956\\
	\hline  $0.810$ & $3.18$ & $0.1160$ & $4.07$ & 0.978\\
	\hline  $0.815$ & $3.20$ & $0.1153$ & $4.03$ & 0.981\\
	\hline  $0.820$ & $3.26$ & $0.1082$ & $3.94$ & 0.982 \\
	\hline  $0.825$ & $3.30$ & $0.1044$ & $3.92$ & 0.991 \\
	\hline  $0.849$ & $3.48$ & $0.0903$ & $3.65$ & 0.999\\
	\hline     
\end{tabular}
\caption{\footnotesize{Table of properties of metastable Q-balls for $\lambda =1$
 and $g = 1$.}}  
\end{center}
\end{table}

\newpage
\section{Conclusions} 

            We have provided examples of inflatonic 
Q-ball solutions of the SUSY 
D-term hybrid inflation scalar field equations for 
typical values of the dimensionless couplings $\lambda$
and $g$, including the special case $\lambda = \sqrt{2}g$,
corresponding to F-term inflation. Since $E/Q_{S} < m_{S}$
for these solutions, $Q_{S}$ conservation implies that 
the Q-balls are stable up to the 
decay of the inflaton sector particles they are made of to 
particles in the minimal SUSY Standard Model sector. 
Inflatonic Q-balls may form at the end of SUSY hybrid inflation 
via the formation
of neutral condensate lumps and their subsequent decay into Q-ball, 
anti-Q-ball pairs. 
In this case there would be a highly inhomogeneous post-inflation era, 
with the energy density of the Universe concentrated inside the Q-balls 
and reheating via Q-ball decay. We hope to
discuss in detail the classically stable Q-ball solutions and the 
process of Q-ball formation following SUSY hybrid inflation in future 
work.

\end{document}